\documentclass[conference,twoside,10pt]{IEEEtran}
\usepackage{spconf,epsfig,cite}
\usepackage{units}
\usepackage{multirow}
\usepackage{amsmath,graphicx,subfig}
\usepackage{algorithm,algorithmic}
\usepackage{listings}
\usepackage{tikz}
\usetikzlibrary{decorations.pathreplacing}
\usetikzlibrary{chains,arrows,shapes,spy}
\usepackage{caption}
\usepackage{enumitem}
\tikzset{font={\fontsize{9}{11.0476pt}\selectfont}}
\usepackage{mathtools,amssymb,cuted}
\usepackage{pgfplots}
\pgfplotsset{compat=newest}
\usepackage{romannum}
\usepackage{boldline}
\usepackage{adjustbox}
\usepackage{caption}
\newcommand{\cev}[1]{\reflectbox{\ensuremath{\vec{\reflectbox{\ensuremath{#1}}}}}}

\definecolor{r}{rgb}{0, 0, 0}
\definecolor{g}{rgb}{0, 0, 0}

\title{ \Large {State Machine-based Waveforms for Channels With 1-Bit Quantization and Oversampling With Time-Instance Zero-Crossing Modulation} }

\name{{Diana~M.~V.~Melo,~Lukas~T.~N.~Landau~and~Rodrigo~C.~de~Lamare}}
\address{Centre for Telecommunications Studies,\\
	Pontifical Catholic University of Rio de Janeiro,\\ 
	Rio de Janeiro, Brazil 22453-900\\
	Email: diana;lukas.landau;delamare@cetuc.puc-rio.br}
	
\begin{document}
\maketitle
\begin{abstract}
Systems with 1-bit quantization and oversampling are promising for the Internet of Things (IoT) devices in order to reduce the power consumption of the analog-to-digital-converters. The novel time-instance zero-crossing (TI ZX) modulation is a promising approach for this kind of channels but existing studies rely on optimization problems with high computational complexity and delay. In this work, we propose a practical waveform design based on the established TI ZX modulation for a multiuser multi-input multi-output (MIMO) downlink scenario with 1-bit quantization and temporal oversampling at the receivers. In this sense, the proposed temporal transmit signals are constructed by concatenating segments of coefficients which convey the information into the time-instances of zero-crossings according to the TI ZX mapping rules. The proposed waveform design is compared with other methods from the literature. The methods are compared in terms of bit error rate and normalized power spectral density. Numerical results show that the proposed technique is suitable for multiuser MIMO system with 1-bit quantization while tolerating some small amount of out-of-band radiation.
\end{abstract}
\section{Introduction}
Future wireless communication technologies are envisioned to support a large number of the Internet of Things (IoT) devices which require to have low power consumption and low complexity.  Low resolution analog-to-digital  converters (ADCs) are suitable to meet the IoT requirements since the power consumption in the ADCs increase exponentially with its amplitude resolution \cite{761034}. 
The loss of information caused by the coarse quantization can be partially compensated by increasing the sampling rate. Employing temporal $M_\mathrm{Rx}$-fold oversampling, rates of $\log_2(M_\mathrm{Rx}+1)$ bits per Nyquist interval are achievable in a noise free environment \cite{shamai_1994}. The authors in \cite{Landau_CL2017} study the maximization of the achievable rate for systems with 1-bit quantization and oversampling in the presence of noise. Other studies that consider systems with 1-bit quantization and oversampling employ ASK transmit sequences \cite{Landau_EURASIP,Son2019} and 16 QAM modulation \cite{Landaus_2013}. Other practical methods are based on the idea presented in \cite{shamai_1994}, where the information is conveyed into the zero-crossings. An example is the study presented in \cite{deng2019bandlimited}, where the waveform is constructed by concatenating sequences which convey the information into the zero-crossings. This study shows that similar data rates to the one presented in \cite{shamai_1994} can be achieved over noisy channels with relatively low  out-of-band radiation. Some other practical methods which convey the information into the zero-crossings include runlength-limited (RLL) sequences \cite{NeuhausICC2020,Neuhaus2020PIMRC}.

The benefits of  1-bit quantization and oversampling have been studied in \cite{ALi_Emil_2020,Shao_Landau2021} for multiple-input multiple-output (MIMO) channels in uplink transmission. Moreover, the studies \cite{LarsonB_2017,Neuhaus_ViVeros} investigate sequences for downlink MIMO systems with 1-bit quantization and oversampling. In this regard, in \cite{LarsonB_2017} it is presented the quantization precoding method which considers as optimization criterion the maximization of the minimum distance to the decision threshold (MMDDT) which was proposed in \cite{Landaus_2013}. The quantization precoding technique relies on an exhaustive codebook search which allows simple Hamming distance detection. Superior precoding schemes for MIMO downlink scenarios have been investigated in \cite{viveros2020_ICASSP,viveros2020_WSA}, where a novel time-instance zero-crossing (TI ZX) modulation is introduced. This novel  modulation follows the idea of \cite{shamai_1994} by allocating the information into the time-instance of zero-crossings in order to reduce the number of zero-crossings of the signal. The study in \cite{viveros2020_ICASSP} relies on a precoding technique based on the MMDDT criterion with spatial zero-forcing (ZF) precoding and TI ZX modulation, whereas \cite{viveros2020_WSA} proposes an optimal temporal-spatial precoding technique with TI ZX modulation along with an minimum mean square error (MMSE) solution. Other studies that consider novel TI ZX modulation schemes have been presented in \cite{Neuhaus_ViVeros,Viveros_asilo_2020,Viveros_ssp} where the computational complexity is reduced \cite{Viveros_asilo_2020}. In \cite{Viveros_ssp} the minimization of the transmit power under quality of service constraint is considered as an objective. The study in \cite{Neuhaus_ViVeros} investigates the spectral efficiency of MIMO systems with sequences constructed with the TI ZX modulation and RLL sequences.

In this work, we propose a TI ZX waveform design for multiuser MIMO downlink channels with 1-bit quantization and oversampling where  a defined level of  out-of-band radiation is tolerated. The proposed waveform design considers the novel TI ZX modulation from \cite{viveros2020_ICASSP,viveros2020_WSA} and follows a similar idea as presented in \cite{deng2019bandlimited}. The proposed method conveys the information into the time-instances of zero-crossings but instead of considering sequences of samples, input bits are mapped into waveform segments according to the TI ZX mapping rules  \cite{viveros2020_ICASSP,viveros2020_WSA}. The temporal precoding vector is then used in conjunction with a simple pulse shaping filter. The optimal set of coefficients is computed with an optimization problem which is formulated to maximize the minimum distance to the decision threshold, constrained with some tolerated out-of-band radiation. Finally, the numerical results are evaluated considering the bit error rate (BER) and the power spectral density (PSD). The proposed waveform design is compared with the transceiver waveform design from \cite{deng2019bandlimited} and the TI ZX MMDDT precoding \cite{viveros2020_ICASSP}. The transceiver waveform design \cite{deng2019bandlimited} was adapted for MIMO channels.  The simulation results show that the proposed waveform design is comparable in terms of BER performance to the one presented for TI ZX MMDDT precoding while having a lower computational complexity since the waveform optimization is done once and is suitable for any input sequence of bits.

The rest of the paper is organized as follows: The system model is introduced in Section~\ref{sec:system_model}. Then,  Section~\ref{sec:TIZX_modulation} describes the novel TI ZX modulation. Section~\ref{waveform_design} explains the proposed waveform design optimization  including the autocorrelation function for TI ZX modulated sequences. The simulation results are provided in Section~\ref{sec:numericalR} and finally, the conclusions are given in Section~\ref{sec:Fconclusions}.

Notation: In the paper all scalar values, vectors and matrices are represented by: ${a}$, ${\boldsymbol{x}}$ and ${\boldsymbol{X}}$, respectively.

\section{System Model}
\label{sec:system_model}
\begin{figure}[!t]
    \begin{center}
    \includegraphics[width=7.8cm]{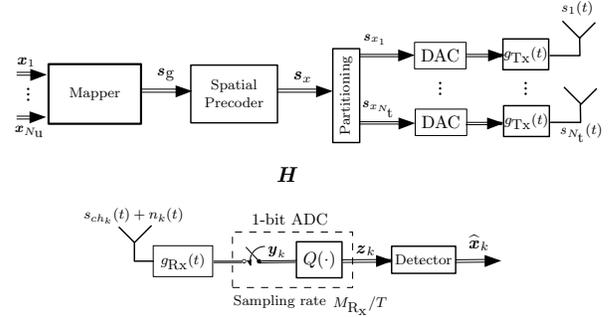}
    \caption{Considered multi-user MIMO downlink system model.}
    \label{fig:system_mimomodel}       
    \end{center}
\end{figure}
In this study, a multiuser MIMO downlink scenario with $N_{\text{u}}$ single antenna users and $N_{\text{t}}$ transmit antennas at the base station (BS), is considered as shown in Fig. 1. Transmission blocks of $N$ symbols ($N$ Nyquist intervals) are considered. The input sequences of symbols ${\boldsymbol{x}_k}$ are mapped  using the TI ZX mapping and the set of coefficients $\mathcal{G}$ which yields the temporal precoding vector $\boldsymbol{s}_{g_{k}} \in \mathbb{C}^{M_\mathrm{Rx}N}$, where $M_\mathrm{Rx} / T$ denotes the sampling rate and $T$ refers to the symbol duration.  Moreover, the transmit filter $g_\mathrm{Tx}(t)$ and receive filter $g_\mathrm{Rx}(t)$ are presented, where the combined waveform is given by $v(t) = \left( g_\mathrm{Tx} * g_\mathrm{Rx} \right)(t)$. Furthermore 1-bit quantization is applied at the receivers. \textcolor{r}{The channel matrix $\boldsymbol{H} \in \mathbb{C}^{N_\mathrm{u} \times N_\mathrm{t}}$ is known at the base station and is considered to be  frequency-flat fading as typically assumed for narrowband IoT systems \cite{LarsonB_2017}}.
Then, with \textcolor{r}{the stacked temporal precoding vector} $\boldsymbol{s}_{g}= \left [ {\boldsymbol{s}}_{g_{1}}^{T}, {\boldsymbol{s}}_{g_{2}}^{T}, \cdots,{\boldsymbol{s}}_{g_{k}}^{T}, \cdots, {\boldsymbol{s}}_{g_{N_{\text{u}}}}^{T} \right ]^{T}$, the received signal $\boldsymbol{z} \in {\mathbb{C}^{N_{\text{tot}}N_{\text{u}}}}$ can be expressed by stacking the received samples of the $N_{\text{u}}$ users as follows:
\textcolor{r}{
\begin{align}
\label{z_receiv}
{\boldsymbol{z}}  
&= 
Q_{1} 
\left (
\left (     {\boldsymbol{H}\boldsymbol{P}_{\mathrm{sp}}} \otimes    \boldsymbol{I}_{N_{\text{tot}}}     \right )
  \left ( \boldsymbol{I}_{N_{\text{u}}} \otimes \boldsymbol{V} \right ) {\boldsymbol{s}}_{\mathrm{g}} + \left ( \boldsymbol{I}_{N_{\text{u}}} \otimes \boldsymbol{G}_{\text{Rx}} \right ) {\boldsymbol{n}} \right ) \notag \\
   &= 
    Q_{1} 
\left (
\left (     {\boldsymbol{H}\boldsymbol{P}_{\mathrm{sp}}} \otimes   \boldsymbol{V}     \right )
{\boldsymbol{s}}_{\mathrm{g}} + \left ( \boldsymbol{I}_{N_{\text{u}}} \otimes \boldsymbol{G}_{\text{Rx}} \right ) {\boldsymbol{n}} \right ) \notag \\
  &= 
Q_{1} 
\left (
{\boldsymbol{H}}_{\text{eff}}   {\boldsymbol{s}}_{\mathrm{g}} +  \boldsymbol{G}_{\text{Rx,eff}}  {\boldsymbol{n}}\right )  
  \text{,}    
\end{align}}
where \textcolor{r}{$Q_{1}(\cdot)$} corresponds the  1-bit quantization operator, ${\boldsymbol{n}} \in \mathbb{C}^{3N_{\text{tot}}N_{\text{u}}}$ denotes a vector with zero-mean complex Gaussian noise samples with variance $\sigma^{2}_{{n}}$ with $N_{\text{tot}} = NM_\mathrm{Rx}$. The waveform matrix $\boldsymbol{V}$ with size $N_{\text{tot}} \times N_{\text{tot}}$ is given by
 \begin{align}
\scriptsize{
\label{eq:MatrixV}
  \mathbb{\boldsymbol{V}} = \;
   \begin{bmatrix}
      v\left ( 0 \right ) & v\left ( \frac{T}{M_{\text{Rx}}} \right ) & \cdots &  v\left (T N \right ) \\
      v\left ( -\frac{T}{M_{\text{Rx}}} \right ) & v\left ( 0 \right ) & \cdots &  v\left (T \left ( N-\frac{1}{M_{\text{Rx}}} \right ) \right ) \\
      \vdots & \vdots & \ddots & \vdots \\
			v\left (-T N \right ) &  v\left (T \left ( -N+\frac{1}{M_{\text{Rx}}} \right ) \right ) & \cdots &  v\left ( 0 \right )
   \end{bmatrix} \text{.}
}
\end{align}
The receive filter $g_\mathrm{Rx}$ is represented in discrete time by the matrix $\boldsymbol{G}_{\textrm{Rx}}$ with size $N_{\text{tot}}  \times 3N_{\text{tot}}$ and is denoted as 
\begin{align}
\label{eq:GRx}
\boldsymbol{G}_{\textrm{Rx}}=  a_{\textrm{Rx}} \begin{bmatrix}
\left[ \ \boldsymbol{g}_{\text{Rx}}^T \ \right]  \ 0 \cdots \ \ \ 0  \\
0 \ \left[ \ \boldsymbol{g}_{\text{Rx}}^T \ \right] \ 0 \cdots 0 \\
\ddots  \ddots \ddots    \\
0 \cdots \ \ \  0 \ \left[ \ \boldsymbol{g}_{\text{Rx}}^T  \ \right]   
\end{bmatrix}
\text{,}
\end{align} 
with
$\boldsymbol{g}_{\text{Rx}} = [ g_{\text{Rx}} (-T  ( N + \frac{1}{M_{\text{Rx}}}  )  ),
g_{\text{Rx}}(-T  ( N + \frac{1}{M_{\text{Rx}}} )  + \frac{T}{M_{\text{Rx}}}  ), \ldots,$
$ g_{\text{Rx}} (T ( N + \frac{1}{M_{\text{Rx}}} )  ) ]^T $ and $a_{\textrm{Rx}}=  (T /  M_{\text{Rx}} )^{1/2} $. 
The matrix $\boldsymbol{P}_{\mathrm{sp}} = c_{\mathrm{zf}}{\boldsymbol{H}}^{\text{H}}\left ( {\boldsymbol{H}}{\boldsymbol{H}}^{\text{H}} \right )^{-1} $ denotes the spatial zero-forcing precoder. The matrix  $\boldsymbol{P}_{\mathrm{sp}}$ is normalized such that the spatial precoder does not change the signal power. As in \cite{viveros2020_ICASSP} the normalization factor $c_{\mathrm{zf}}$ is given by
\begin{align}
c_{\text{zf}} =  \left (  N_{\text{u}}/ \mathrm{trace} \left (  \left ( {\boldsymbol{H}}{\boldsymbol{H}}^{\text{H}} \right )^{-1} \right )  \right )^{\frac{1}{2}} \text{.}
\end{align}

\section{Time-Instance Zero-Crossing Mapping}
\label{sec:TIZX_modulation}
The TI ZX modulation was  proposed  in the studies \cite{viveros2020_ICASSP} and \cite{viveros2020_WSA}  for systems with 1-bit quantization and oversampling. The TI ZX modulation conveys the information into the time-instances of zero-crossings and also considers the absence of zero-crossing during a symbol interval as a valid symbol, \textcolor{g}{different from \cite{shamai_1994} and \cite{deng2019bandlimited}}. 
To build the mapped sequence, each symbol  ${x}_{i}$ drawn from the set $  \mathcal{X}_{\text{in}}:=  \left \{ b_{1},b_{2},
 \cdots , b_{R_{\text{in}}}\right \}$ with $R_{\text{in}} = M_{\text{Rx}} + 1$,  is mapped into a binary codeword ${\boldsymbol{c}}_{\textrm{s}_i}$ with $M_{\text{Rx}}$ samples. As mentioned, one of the possible symbols corresponds to the pattern that does not contain a zero-crossing. The mapping depends on the last sample of the previous symbol interval, namely $\rho \in \{1,-1\}$. Hence, the TI ZX mapping provides two possible codewords ${\boldsymbol{c}}_{\textrm{s}_i}$ for each valid symbol ${x}_{i}$ which convey the same zero-crossing information. Then, for coding and decoding  of the first transmit symbol, a pilot sample $\rho_{\textrm{b}} \in \{1,-1\}$  is required.

\section{Waveform Design Optimization}
\label{waveform_design}
The proposed waveform design, suitable for systems with 1-bit quantization and oversampling,  considers the novel TI ZX modulation \cite{viveros2020_ICASSP,viveros2020_WSA}, in conjunction with the optimization of a set of coefficients. The  proposed waveform is built by concatenating segment sequences, i.e., subsequences, described by the coefficients which contain zero-crossings at the desired time-instances. 
The proposed waveform design relies on the transmit  and receive filters $g_\mathrm{Tx}(t)$ and  $g_\mathrm{Rx}(t)$ which preserve the zero-crossing time-instance. Different to prior studies \cite{viveros2020_ICASSP,viveros2020_WSA}, the sequence is no  longer binary but is defined by the set of coefficients $\mathcal{G}$, so that each symbol ${x}_{i}$ drawn from the set $  \mathcal{X}_{\text{in}}$  is mapped into a codeword $\boldsymbol{g}_{i}$ with $M_{\text{Rx}}$ different coefficients which convey the information into the time-instances of zero-crossings. \textcolor{g}{Moreover, it is considered that sequences are constructed for real and imaginary parts independently. In the following, a real values process is described.} The set of coefficients $\mathcal{G}$ is defined in terms of $\mathcal{G} = \{ \boldsymbol{G}_{+}; \boldsymbol{G}_{-}\}$ where $\boldsymbol{G}_{-} = - \boldsymbol{G}_{+}$, such that they both convey the same zero-crossing information and the sign information of the coefficients depends on the last sample of the previous interval termed $\rho$. Considering bit sequences as input and the Gray coding for TI ZX modulation shown in \cite[Table~II]{viveros2020_ICASSP}, $n_{\text{s}}=2\varrho$ different states  can be defined. In this context, the set $\boldsymbol{G} = \left [ \boldsymbol{g}_{1}^T;\boldsymbol{g}_{2}^T;\cdots;\boldsymbol{g}_{\varrho}^T \right ]$ is presented, where $\boldsymbol{g}_{i} = \left [ g_{i,1},g_{i,2},\cdots g_{i,q}\right ] $ and $\rho = \textrm{sgn} \left ( g_{i,M_{\text{Rx}}} \right )$. Then, as initially established, the symbol $x_{i}$ is mapped in the segment $\boldsymbol{g}_{i}$. The pilot sample $\rho_{\text{b}}$ is required for the encoding and decoding processes of the first symbol $x_{1}$. Finally, the input sequence of symbols $\boldsymbol{x}_{k}$ is mapped in the sequence $\boldsymbol{s}_{g_{k}}$ with length $N_{\text{tot}}$ by concatenating all the segments $\boldsymbol{g}_{i}$ such that, $\boldsymbol{s}_{g_{k}}=[\boldsymbol{g}_{0}^T,\ldots,\boldsymbol{g}_{N-1}^T
]^T$. Note that the pilot sample  $\rho_{\text{b}}$ is predefined and known at the receivers, hence not included in the precoding vector $\boldsymbol{s}_{g_{k}}$.
\subsection{Autocorrelation for TI ZX Modulation}
In this section, it is described how  to compute the autocorrelation function of the TI ZX modulated signal, considering the set of coefficients $\mathcal{G}$ which conveys the information into the time-instances of zero-crossings. 

To obtain the autocorrelation function, the TI ZX modulation system is converted to a finite-state machine where the current output values are determined only by its current state which corresponds to  an equivalent Moore machine \cite{Peter2021}. For $M_{\text{Rx}} =3$,  one symbol in terms of two bits  is mapped in one output pattern, so  $\varrho = 4$ and $n_\text{s} = 8$ different states are presented. While for $M_{\text{Rx}} = 2$ sequences of symbols are considered in terms of mapping  three bits segments \textcolor{r}{in} four samples, such that $\varrho = 8$ with $n_\text{s} = 16$ different states. 
Table~\ref{tab:moore3} and Table~\ref{tab:moore2} provide the equivalent Moore machine for $M_{\text{Rx}} = 3$ and $M_{\text{Rx}}= 2$, respectively. The states  with positive subscripts  represent sequences for $\rho=1$ and states with negative subscripts represent sequences for $\rho=-1$. 

Considering a symmetric machine there are $m = \varrho M_{\text{Rx}} = 12$ different coefficients for $M_{\text{Rx}} = 3$. On the other hand, for $M_{\text{Rx}} = 2$  sequences of symbols are considered such that there are $m = 2\varrho M_{\text{Rx}} = 32$ different coefficients.

The state transition probability  
matrix $\boldsymbol{Q}$ of the equivalent Moore machine, with dimensions $n_\text{s} \times n_\text{s}$ is defined for  i.i.d. input bits, all valid state transitions have equal probability $p$ with $p=1 /4$ for  $M_{\text{Rx}} = 3$ and $p=1 /8$ for  $M_{\text{Rx}} = 2$. Furthermore, the vector $\boldsymbol{\pi} = (1/n_\text{s}) \boldsymbol{1}$ of length $n_\text{s}$  corresponds to the  stationary distribution of the equivalent Moore machine, which implies $\boldsymbol{\pi}^{T}\boldsymbol{Q} = \boldsymbol{\pi}^{T}$. Then, the matrix $\boldsymbol{\Gamma}$ with dimensions $n_\text{s} \times M_{\text{Rx}} $ for  $M_{\text{Rx}} = 3$ and $n_\text{s} \times 2M_{\text{Rx}} $ for  $M_{\text{Rx}} = 2$  is defined which contains  the Moore machine’s output $\boldsymbol{g}_{i}$. The  block-wise correlation matrix of the TI ZX mapping output is given by \cite[eq.~3.46]{Immink}
\begin{align}
    \boldsymbol{R}^{\kappa}_{\boldsymbol{g}}=  \mathrm{E}  \{{\boldsymbol{g}}_{{\kappa'}}{\boldsymbol{g}}_{{\kappa'+ \kappa}}^T \}  = \boldsymbol{\Gamma}^{T}\boldsymbol{\Pi} \boldsymbol{Q}^{\left | \kappa \right |}\boldsymbol{\Gamma} \text{.}
\end{align}
Then, the average autocorrelation function
$\boldsymbol{r}_{g}$ of the TI ZX modulation output sequence  can be obtained as \cite[eq.~3.39]{Immink}
\begin{align}
\label{autocorrAverage}
    \boldsymbol{r}_{\boldsymbol{g}}[kq+l]=  \frac{1}{q}\left ( \sum_{i=1}^{q-l}\left [ \boldsymbol{R}_{\boldsymbol{g}}^{k} \right ]_{i,l+i} + \sum_{i=q-l+1}^{q}\left [ \boldsymbol{R}_{\boldsymbol{g}}^{k+1} \right ]_{i,l+i-q} \right )\text{,}
\end{align}
for $k\in \mathbb{Z}$, $0\leq l\leq q-1$.
\begin{table}[H]
\caption{Equivalent Moore machine for TI ZX mapping for $M_{\text{Rx}} = 3$} 
\label{tab:moore3} \vspace{0pt}
\footnotesize
\begin{center}
\scalebox{0.65}{
\begin{tabular}{|ll|llll|l|}
\hline
\multicolumn{2}{|c|}{\multirow{2}{*}{\begin{tabular}[c]{@{}c@{}}Current \\ state\end{tabular}}} & \multicolumn{4}{c|}{next state}                                                              & \multicolumn{1}{c|}{\multirow{2}{*}{\begin{tabular}[c]{@{}c@{}}output\\ $\boldsymbol{g}_{{i}}$\end{tabular}}} \\ \cline{3-6}
\multicolumn{2}{|c|}{}                                                                          & \multicolumn{1}{l|}{00}    & \multicolumn{1}{l|}{01}    & \multicolumn{1}{l|}{11}    & 10    & \multicolumn{1}{c|}{}                                                                                         \\ \hline
\multicolumn{2}{|l|}{$1_+$}                                                                     & \multicolumn{1}{l|}{$1_+$} & \multicolumn{1}{l|}{$2_+$} & \multicolumn{1}{l|}{$3_+$} & $4_+$ & $\;g_{1,1}  \quad   \;g_{1,2}    \quad   \;g_{1,3}$                                                           \\ \hline
\multicolumn{2}{|l|}{$2_+$}                                                                     & \multicolumn{1}{l|}{$1_-$} & \multicolumn{1}{l|}{$2_-$} & \multicolumn{1}{l|}{$3_-$} & $4_-$ & $\;g_{2,1}  \quad   \;g_{2,2}            \;-g_{2,3}$                                                          \\ \hline
\multicolumn{2}{|l|}{$3_+$}                                                                     & \multicolumn{1}{l|}{$1_-$} & \multicolumn{1}{l|}{$2_-$} & \multicolumn{1}{l|}{$3_-$} & $4_-$ & $\;g_{3,1}          \;-g_{3,2}          \;-g_{3,3}$                                                           \\ \hline
\multicolumn{2}{|l|}{$4_+$}                                                                     & \multicolumn{1}{l|}{$1_-$} & \multicolumn{1}{l|}{$2_-$} & \multicolumn{1}{l|}{$3_-$} & $4_-$ & $-g_{4,1}          \;-g_{4,2}           \;-g_{4,3}$                                                           \\ \hline
\multicolumn{2}{|l|}{$4_+$}                                                                     & \multicolumn{1}{l|}{$1_-$} & \multicolumn{1}{l|}{$2_-$} & \multicolumn{1}{l|}{$3_-$} & $4_-$ & $-g_{4,1}          \;-g_{4,2}           \;-g_{4,3}$                                                           \\ \hline
\multicolumn{2}{|l|}{$1_-$}                                                                     & \multicolumn{1}{l|}{$1_-$} & \multicolumn{1}{l|}{$2_-$} & \multicolumn{1}{l|}{$3_-$} & $4_-$ & $-g_{1,1}             \;-g_{1,2}             \;-g_{1,3}$                                                           \\ \hline
\multicolumn{2}{|l|}{$2_-$}                                                                     & \multicolumn{1}{l|}{$1_+$} & \multicolumn{1}{l|}{$2_+$} & \multicolumn{1}{l|}{$3_+$} & $4_+$ & $-g_{2,1}             \;-g_{2,2}    \quad    \;g_{2,3}$                                                           \\ \hline
\multicolumn{2}{|l|}{$3_-$}                                                                     & \multicolumn{1}{l|}{$1_+$} & \multicolumn{1}{l|}{$2_+$} & \multicolumn{1}{l|}{$3_+$} & $4_+$ & $-g_{3,1}   \quad     \;g_{3,2}     \quad    \;g_{3,3}$                                                           \\ \hline
\multicolumn{2}{|l|}{$4_-$}                                                                     & \multicolumn{1}{l|}{$1_+$} & \multicolumn{1}{l|}{$2_+$} & \multicolumn{1}{l|}{$3_+$} & $4_+$ & $\;g_{4,1}  \quad      \;g_{4,2}    \quad    \;g_{4,3}$                                                           \\ \hline
\end{tabular}}
\end{center}
 \vspace{-2mm}
\end{table}
\begin{table}[H]
\caption{Equivalent Moore machine for TI ZX mapping for $M_{\text{Rx}} = 2$} 
\label{tab:moore2} \vspace{0pt}
\footnotesize
\begin{center}
\scalebox{0.65}{
\begin{tabular}{|cl|llllllll|c|}
\hline
\multicolumn{2}{|c|}{\multirow{2}{*}{\begin{tabular}[c]{@{}c@{}}Current \\ state\end{tabular}}} & \multicolumn{8}{c|}{next state}                                                                                                                                                                                  & \multirow{2}{*}{\begin{tabular}[c]{@{}c@{}}output\\ $\boldsymbol{g}_{{i}}$\end{tabular}}     \\ \cline{3-10}
\multicolumn{2}{|c|}{}                                                                          & \multicolumn{1}{l|}{000}   & \multicolumn{1}{l|}{001}   & \multicolumn{1}{l|}{011}   & \multicolumn{1}{l|}{010}   & \multicolumn{1}{l|}{110}   & \multicolumn{1}{l|}{111}   & \multicolumn{1}{l|}{101}   & 100   &                                                                                              \\ \hline
\multicolumn{2}{|l|}{$1_+$}                                                                     & \multicolumn{1}{l|}{$1_+$} & \multicolumn{1}{l|}{$2_+$} & \multicolumn{1}{l|}{$3_+$} & \multicolumn{1}{l|}{$4_+$} & \multicolumn{1}{l|}{$5_+$} & \multicolumn{1}{l|}{$6_+$} & \multicolumn{1}{l|}{$7_+$} & $8_+$ & \multicolumn{1}{l|}{$\;g_{1,1}  \quad   \;g_{1,2}    \quad   \;g_{1,3}  \quad \; g_{1,4}  $} \\ \hline
\multicolumn{2}{|l|}{$2_+$}                                                                     & \multicolumn{1}{l|}{$1_-$} & \multicolumn{1}{l|}{$2_-$} & \multicolumn{1}{l|}{$3_-$} & \multicolumn{1}{l|}{$4_-$} & \multicolumn{1}{l|}{$5_-$} & \multicolumn{1}{l|}{$6_-$} & \multicolumn{1}{l|}{$7_-$} & $8_-$ & \multicolumn{1}{l|}{$\;g_{2,1}  \quad   \;g_{2,2}    \quad   \;g_{2,3}  \;- g_{2,4}  $} \\ \hline
\multicolumn{2}{|l|}{$3_+$}                                                                     & \multicolumn{1}{l|}{$1_-$} & \multicolumn{1}{l|}{$2_-$} & \multicolumn{1}{l|}{$3_-$} & \multicolumn{1}{l|}{$4_-$} & \multicolumn{1}{l|}{$5_-$} & \multicolumn{1}{l|}{$6_-$} & \multicolumn{1}{l|}{$7_-$} & $8_-$ & \multicolumn{1}{l|}{$\;g_{3,1}  \quad   \;g_{3,2}       \;-g_{3,3}  \;- g_{3,4}  $} \\ \hline
\multicolumn{2}{|l|}{$4_+$}                                                                     & \multicolumn{1}{l|}{$1_-$} & \multicolumn{1}{l|}{$2_-$} & \multicolumn{1}{l|}{$3_-$} & \multicolumn{1}{l|}{$4_-$} & \multicolumn{1}{l|}{$5_-$} & \multicolumn{1}{l|}{$6_-$} & \multicolumn{1}{l|}{$7_-$} & $8_-$ & \multicolumn{1}{l|}{$\;g_{4,1}    \;-g_{4,2}       \;-g_{4,3}  \;- g_{4,4}  $} \\ \hline
\multicolumn{2}{|l|}{$5_+$}                                                                     & \multicolumn{1}{l|}{$1_+$} & \multicolumn{1}{l|}{$2_+$} & \multicolumn{1}{l|}{$3_+$} & \multicolumn{1}{l|}{$4_+$} & \multicolumn{1}{l|}{$5_+$} & \multicolumn{1}{l|}{$6_+$} & \multicolumn{1}{l|}{$7_+$} & $8_+$ & \multicolumn{1}{l|}{$\;g_{5,1}    \;-g_{5,2}       \;-g_{5,3}   \quad  \; g_{5,4}  $} \\ \hline
\multicolumn{2}{|l|}{$6_+$}                                                                     & \multicolumn{1}{l|}{$1_+$} & \multicolumn{1}{l|}{$2_+$} & \multicolumn{1}{l|}{$3_+$} & \multicolumn{1}{l|}{$4_+$} & \multicolumn{1}{l|}{$5_+$} & \multicolumn{1}{l|}{$6_+$} & \multicolumn{1}{l|}{$7_+$} & $8_+$ & \multicolumn{1}{l|}{$-g_{6,1}    \;-g_{6,2}       \;-g_{6,3}   \quad  \; g_{6,4}  $} \\ \hline
\multicolumn{2}{|l|}{$7_+$}                                                                     & \multicolumn{1}{l|}{$1_-$} & \multicolumn{1}{l|}{$2_-$} & \multicolumn{1}{l|}{$3_-$} & \multicolumn{1}{l|}{$4_-$} & \multicolumn{1}{l|}{$5_-$} & \multicolumn{1}{l|}{$6_-$} & \multicolumn{1}{l|}{$7_-$} & $8_-$ & \multicolumn{1}{l|}{$-g_{7,1}    \;-g_{7,2}       \;-g_{7,3}    \;- g_{7,4}  $} \\ \hline
\multicolumn{2}{|l|}{$8_+$}                                                                     & \multicolumn{1}{l|}{$1_+$} & \multicolumn{1}{l|}{$2_+$} & \multicolumn{1}{l|}{$3_+$} & \multicolumn{1}{l|}{$4_+$} & \multicolumn{1}{l|}{$5_+$} & \multicolumn{1}{l|}{$6_+$} & \multicolumn{1}{l|}{$7_+$} & $8_+$ & \multicolumn{1}{l|}{$-g_{8,1}    \;-g_{8,2}     \quad  \;g_{8,3}   \quad  \; g_{8,4}  $} \\ \hline
\multicolumn{2}{|l|}{$1_-$}                                                                     & \multicolumn{1}{l|}{$1_-$} & \multicolumn{1}{l|}{$2_-$} & \multicolumn{1}{l|}{$3_-$} & \multicolumn{1}{l|}{$4_-$} & \multicolumn{1}{l|}{$5_-$} & \multicolumn{1}{l|}{$6_-$} & \multicolumn{1}{l|}{$7_-$} & $8_-$ & \multicolumn{1}{l|}{$-g_{1,1}    \;-g_{1,2}       \;-g_{1,3}    \;- g_{1,4}  $} \\ \hline
\multicolumn{2}{|l|}{$2_-$}                                                                     & \multicolumn{1}{l|}{$1_+$} & \multicolumn{1}{l|}{$2_+$} & \multicolumn{1}{l|}{$3_+$} & \multicolumn{1}{l|}{$4_+$} & \multicolumn{1}{l|}{$5_+$} & \multicolumn{1}{l|}{$6_+$} & \multicolumn{1}{l|}{$7_+$} & $8_+$ & \multicolumn{1}{l|}{$-g_{2,1}    \;-g_{2,2}       \;-g_{2,3}   \quad  \; g_{2,4}  $} \\ \hline
\multicolumn{2}{|l|}{$3_-$}                                                                     & \multicolumn{1}{l|}{$1_+$} & \multicolumn{1}{l|}{$2_+$} & \multicolumn{1}{l|}{$3_+$} & \multicolumn{1}{l|}{$4_+$} & \multicolumn{1}{l|}{$5_+$} & \multicolumn{1}{l|}{$6_+$} & \multicolumn{1}{l|}{$7_+$} & $8_+$ & \multicolumn{1}{l|}{$-g_{3,1}    \;-g_{3,2}     \quad  \;g_{3,3}   \quad  \; g_{3,4}  $} \\ \hline
\multicolumn{2}{|l|}{$4_-$}                                                                     & \multicolumn{1}{l|}{$1_+$} & \multicolumn{1}{l|}{$2_+$} & \multicolumn{1}{l|}{$3_+$} & \multicolumn{1}{l|}{$4_+$} & \multicolumn{1}{l|}{$5_+$} & \multicolumn{1}{l|}{$6_+$} & \multicolumn{1}{l|}{$7_+$} & $8_+$ & \multicolumn{1}{l|}{$-g_{4,1}  \quad   \;g_{4,2}    \quad   \;g_{4,3}  \quad \; g_{4,4}  $} \\ \hline
\multicolumn{2}{|l|}{$5_-$}                                                                     & \multicolumn{1}{l|}{$1_-$} & \multicolumn{1}{l|}{$2_-$} & \multicolumn{1}{l|}{$3_-$} & \multicolumn{1}{l|}{$4_-$} & \multicolumn{1}{l|}{$5_-$} & \multicolumn{1}{l|}{$6_-$} & \multicolumn{1}{l|}{$7_-$} & $8_-$ & \multicolumn{1}{l|}{$-g_{5,1}  \quad   \;g_{5,2}    \quad   \;g_{5,3}  \;- g_{5,4}  $} \\ \hline
\multicolumn{2}{|l|}{$6_-$}                                                                     & \multicolumn{1}{l|}{$1_-$} & \multicolumn{1}{l|}{$2_-$} & \multicolumn{1}{l|}{$3_-$} & \multicolumn{1}{l|}{$4_-$} & \multicolumn{1}{l|}{$5_-$} & \multicolumn{1}{l|}{$6_-$} & \multicolumn{1}{l|}{$7_-$} & $8_-$ & \multicolumn{1}{l|}{$\;g_{6,1}  \quad   \;g_{6,2}    \quad   \;g_{6,3}  \;- g_{6,4}  $} \\ \hline
\multicolumn{2}{|l|}{$7_-$}                                                                     & \multicolumn{1}{l|}{$1_+$} & \multicolumn{1}{l|}{$2_+$} & \multicolumn{1}{l|}{$3_+$} & \multicolumn{1}{l|}{$4_+$} & \multicolumn{1}{l|}{$5_+$} & \multicolumn{1}{l|}{$6_+$} & \multicolumn{1}{l|}{$7_+$} & $8_+$ & \multicolumn{1}{l|}{$\;g_{7,1}  \quad   \;g_{7,2}    \quad   \;g_{7,3}  \quad \; g_{7,4}  $} \\ \hline
\multicolumn{2}{|l|}{$8_-$}                                                                     & \multicolumn{1}{l|}{$1_-$} & \multicolumn{1}{l|}{$2_-$} & \multicolumn{1}{l|}{$3_-$} & \multicolumn{1}{l|}{$4_-$} & \multicolumn{1}{l|}{$5_-$} & \multicolumn{1}{l|}{$6_-$} & \multicolumn{1}{l|}{$7_-$} & $8_-$ & \multicolumn{1}{l|}{$\;g_{8,1}  \quad   \;g_{8,2}       \;-g_{8,3}  \;- g_{8,4}  $} \\ \hline
\end{tabular}}
\end{center}
\vspace{-3mm}
\end{table}

\subsection{Waveform Design}
 \textcolor{r}{For a given set $\mathcal{G}$,  the  autocorrelation function is calculated  with \eqref{autocorrAverage}. With this, the PSD is calculated by}
\begin{align}
\label{psdEquation}
    S(f)= S_{x}(f)\left | G_{\textrm{Tx}}(f)\right |^2 \text{,}
\end{align}
where $G_{\textrm{Tx}}(f)$ refers to the transfer function of the transmit filter $g_{\textrm{Tx}}$ and $S_{x}(f)$ to the PSD  of the transmit sequence
\begin{align}
    S_{x}(f)= \frac{M_{\text{Rx}}}{T}\sum_{l=-\infty }^{\infty}c_{l}e^{j2\pi\frac{lT}{M_{\text{Rx}}}f}\text{,}
\end{align}
where $c_{l}$ denotes the $l$-th element of the  autocorrelation function  from \eqref{autocorrAverage}. By defining a critical frequency $f_{c}$ and a power containment factor $\eta$, the inband power is defined as
\begin{align}
    \int_{-f_{c}}^{f_{c}}S(f)\textrm{df} = \eta P\text{,}
\end{align}
where $P = \int_{-\infty}^{\infty}S(f)\textrm{df}$. \textcolor{g}{Then, 
when considering $g_{\textrm{Rx}}(t)$ and $ g_{\textrm{Tx}}(t)$ as rectangular filters defined as 
\begin{align}
\label{filters}
    g_{\textrm{Rx}}(t) = g_{\textrm{Tx}}(t) =  \sqrt{\frac{1}{\nicefrac{T}{M_\mathrm{Rx}}}}\textrm{rect}\left ( \frac{t}{\nicefrac{T}{M_\mathrm{Rx}}} \right )\text{,}
\end{align}
which yields $\boldsymbol{V}$ as an identity matrix. Then, a non convex constrained optimization problem  which maximizes the minimum distance to the decision threshold $\gamma$ can be formulated as}: 
\begin{equation}
\label{eq:OPproblem}
\begin{aligned}
& \mathrm{minimize}_{\boldsymbol{g}_{u}}
& & -\gamma \\
& \text{subject to}
& &\boldsymbol{g}_{u}  \succ  \gamma\boldsymbol{1} \\
&&& \left \| \boldsymbol{g}_{u} \right \|_{2}^2 \leq m\frac{E_{\text{0}}}{2 N_{\text{tot}}}  \\
&&& \eta(\boldsymbol{g}_{u},f_{c}) \geq  0.95 \text{.}
\end{aligned}
\end{equation}
\textcolor{r}{In contrast to existing methods  \cite{LarsonB_2017,viveros2020_ICASSP}, the optimization process is done only once at the BS regardless of the channel and input sequence. Therefore, the optimization process can be done offline by applying an exhaustive search. When the optimal set of coefficients $\mathcal{G}$ is obtained, the sequence $\boldsymbol{s}_{g_{k}}$ is constructed for each user}. \textcolor{g}{Finally, the average total power   of the complex transmit signal $\boldsymbol{s}_{g}$ is given by}
\begin{align}
\mathbb{E} \left \{\boldsymbol{s}_{g}^H   \boldsymbol{A}^H
\boldsymbol{A}\boldsymbol{s}_{\text{g}}\right \}
=  E_{0}\text{,} 
\end{align}
where $\boldsymbol{A}=
 \boldsymbol{I}_{N_{\text{u}}} \otimes \boldsymbol{G}^T_{\textrm{Tx}}$ and $\boldsymbol{G}_{\textrm{Tx}}$ denotes a Toeplitz matrix of size $N_{\text{tot}}  \times 3N_{\text{tot}} $, which is given by
\begin{align}
    \label{eq:GTx}
    \boldsymbol{G}_{\textrm{Tx}} = a_{\textrm{Tx}} \begin{bmatrix}
        \left[ \ \boldsymbol{g}_{\text{Tx}}^T \ \right]  \ 0 \cdots \ \ \ 0  \\
        0 \ \left[ \ \boldsymbol{g}_{\text{Tx}}^T \ \right] \ 0 \cdots 0 \\
        \ddots  \ddots \ddots    \\
        0 \cdots \ \ \  0 \ \left[ \ \boldsymbol{g}_{\text{Tx}}^T  \ \right]   
    \end{bmatrix},
\end{align} 
with $a_\mathrm{Tx} =  (T / M_{\text{Tx}} )^{1/2} $ and
$\boldsymbol{g}_{\text{Tx}} = \left[ g_{\text{Tx}} (-T  ( N + M_\mathrm{Tx}^{-1}  )  ),\right.$ $\left.g_{\text{Tx}}(-T  ( N + M_\mathrm{Tx}^{-1} )  + T \, M_\mathrm{Tx}^{-1}  ),\ldots, g_{\text{Tx}} (T ( N + M_\mathrm{Tx}^{-1} ) ) \right]^T$\text{.} \textcolor{g}{Note that under the assumption in \eqref{filters}, $\boldsymbol{A}^H
\boldsymbol{A}$ corresponds to the identity matrix of dimensions $N_{\text{tot}}  \times N_{\text{tot}} $. }
\subsection{Detection}
The detection process for the proposed waveform, follows the same process as  for the existing TI ZX waveforms which aims for a low complexity receiver \cite{viveros2020_ICASSP,viveros2020_WSA}. The detection process is done in the same way and separately for each user stream. From the sequence received in \eqref{z_receiv} the corresponding $\boldsymbol{z}_k$ sequences of each user are obtained.
The sequence $\boldsymbol{z}_k$ is segmented into subsequences $\boldsymbol{z}_{\textrm{b}_i}= [ \rho_{i-1}, \boldsymbol{z}_i ]^T \in \{+1,-1\}^{M_\mathrm{Rx}+1}$, where $\rho_{i-1}$ corresponds to the last sample of $\boldsymbol{z}_{\textrm{b}_{i-1}}$ which corresponds to the received sequence of the $(i-1)$ symbol interval. Then the backward mapping process is define such that $\cev{d}: \boldsymbol{z}_{\textrm{b}_i} \rightarrow [\rho_{i-1}, \boldsymbol{c}_{\textrm{s}i}^T]$ \cite{viveros2020_ICASSP,viveros2020_WSA}. In the noise free case it is possible to decode the sequence with the backward mapping process $\cev{d}(\cdot)$. However, in the presence of noise, invalid sequences $\boldsymbol{z}_{\textrm{b}_i}$ may arise that are not possible to \textcolor{g}{detect via  $\cev{d}(\cdot)$}. Hence, the Hamming distance metric is required \cite{LarsonB_2017}  which is defined as
\begin{equation}
    \hat{\boldsymbol{x}}_i = \cev{d}(\boldsymbol{c}), \text{~~with~~} \boldsymbol{c} = \arg \!\!\!\! \min_{\boldsymbol{c}_\mathrm{map} \in \mathcal{M}} \!\! \mathrm{Hamming}(\boldsymbol{z}_{\textrm{b}_i}, \boldsymbol{c}_\mathrm{map}),
\end{equation}
where $\text{Hamming} \; ( \boldsymbol{z}_{\textrm{b}_i},\boldsymbol{c}_{\text{map}}) = \sum_{n=1}^{M_{\text{Rx}}+1}\frac{1}{2}\left | \boldsymbol{z}_{\textrm{b}_{i,n}}- \boldsymbol{c}_{\text{map},n} \right |$ and  $\boldsymbol{c}_\mathrm{map} = [\rho_{i-1},\boldsymbol{c}_{\textrm{s}_i}]^T$, and $\mathcal{M}$ denotes all valid forward mapping codewords.
The detection of the first symbol in the sequence, considers the sample $\rho_{\textrm{b}}$ which then enables the detection process. The real and the imaginary parts are detected independently in separate processes. 
\begin{table}[!t]
    \tiny
    \caption{Simulation parameters}
    \label{tab:num_res_parameters} 
    \begin{center}
\begin{tabular}{|l|c|l|l|c|c|}
\hline
\multicolumn{1}{|c|}{Method}                                             & \multicolumn{1}{l|}{$M_{\text{Rx}}$} & \multicolumn{1}{c|}{Transmit Filter}                      & \multicolumn{1}{c|}{Receive filter}                       & \multicolumn{1}{l|}{$I_{\textrm{b}}$} & \multicolumn{1}{l|}{$O_{\textrm{s}}$} \\ \hline
\multirow{2}{*}{TI ZX MMDDT \cite{viveros2020_ICASSP}} & 2                                    & \multirow{2}{*}{RC  $\alpha = 0.22$}                      & \multirow{2}{*}{RRC $\alpha = 0.22$}                      & 45                                    & 61                                    \\ \cline{2-2} \cline{5-6} 
                                                                         & 3                                    &                                                           &                                                           & 60                                    & 91                                    \\ \hline
ZX transceiver design \cite{deng2019bandlimited}        & 3                                    & RC window $\alpha = 0.1$                                  & Integrate-and-dump                                        & 180                                   & 270                                   \\ \hline
\multirow{2}{*}{TI ZX waveform design}                                   & 2                                    & \multirow{2}{*}{Integrator $\nicefrac{T}{M_\mathrm{Rx}}$} & \multirow{2}{*}{Integrator $\nicefrac{T}{M_\mathrm{Rx}}$} & 45                                    & 60                                    \\ \cline{2-2} \cline{5-6} 
                                                                         & 3                                    &                                                           &                                                           & 60                                    & 90                                    \\ \hline
\end{tabular}
    \end{center}
    \vspace{-6mm}
\end{table}

\begin{figure*}
    \centering
    \begin{minipage}[t]{0.33\textwidth}
      \centering
      \usetikzlibrary{positioning,calc}

\definecolor{mycolor1}{rgb}{0.00000,1.00000,1.00000}%
\definecolor{mycolor2}{rgb}{1.00000,0.00000,1.00000}%
\definecolor{mycolor3}{rgb}{0.83,0.69,0.22}%

\pgfplotsset{every axis label/.append style={font=\footnotesize},
every tick label/.append style={font=\footnotesize}
}


\begin{tikzpicture}[font=\footnotesize] 

\begin{axis}[%
width=2.4in,
height=2.0in,
ymode=log,
xmin  = -10, 
xmax  = 30,
xlabel= {SNR [dB]},
xmajorgrids,
ymin=0.0001,
ymax=1,
ylabel={BER},
ymajorgrids,
legend entries={$M_{\text{Rx}}= 2$, $M_{\text{Rx}} = 3$},
legend style={at={(0,0)},anchor=south  west,draw=black,fill=white,legend cell align=left,font=\tiny}
]
\addlegendimage{smooth,color=red,  densely dashed , line width=1.1pt, every mark/.append style={solid, fill=gray!50}, mark=none,
y filter/.code={\pgfmathparse{\pgfmathresult-0}\pgfmathresult}}
\addlegendimage{smooth,color=blue,  line width=1.1pt, every mark/.append style={solid, fill=gray!50}, mark=none,
y filter/.code={\pgfmathparse{\pgfmathresult-0}\pgfmathresult}}


\addplot+[densely dashed, color=red, every mark/.append style={solid, fill=red!50}, mark=none, mark repeat=1, line width=1.3pt]
  table[row sep=crcr]{%
-10	0.458272222222222\\
-8	0.434516666666667\\
-6	0.40625\\
-4	0.371527777777778\\
-2	0.320388888888889\\
0	0.259522222222222\\
2	0.191277777777778\\
4	0.124283333333333\\
6	0.0655277777777778\\
8	0.0274888888888889\\
10	0.00698888888888889\\
12	0.0014\\
14	0.0001\\
16	0\\
18	0\\
20	0\\
22	0\\
24	0\\
26	0\\
28	0\\
30	0\\
};

\addplot+[solid, color=blue, every mark/.append style={solid, fill=blue!50}, mark=none, mark repeat=1, line width=1.3pt]
  table[row sep=crcr]{%
-10	0.463776666666667\\
-8	0.444889166666667\\
-6	0.419326666666667\\
-4	0.384619166666667\\
-2	0.342468333333333\\
0	0.292285833333333\\
2	0.23956\\
4	0.189190833333333\\
6	0.1420225\\
8	0.0996891666666667\\
10	0.0627008333333333\\
12	0.0326841666666667\\
14	0.0129175\\
16	0.00342583333333333\\
18	0.000538333333333333\\
20	4.08333333333333e-05\\
22	2.5e-06\\
24	0\\
26	0\\
28	0\\
30	0\\
};

\end{axis}

\end{tikzpicture}%
    \end{minipage}%
    \hspace{1mm}%
    \begin{minipage}[t]{0.33\textwidth}
      \centering
      \usetikzlibrary{positioning,calc}

\definecolor{mycolor1}{rgb}{0.00000,1.00000,1.00000}%
\definecolor{mycolor2}{rgb}{1.00000,0.00000,1.00000}%
\definecolor{mycolor3}{rgb}{0.83,0.69,0.22}%

\pgfplotsset{every axis label/.append style={font=\footnotesize},
every tick label/.append style={font=\footnotesize}
}


\begin{tikzpicture}[font=\footnotesize] 

\begin{axis}[%
width=2.4in,
height=2.0in,
ymode=log,
xmin  = -10, 
xmax  = 30,
xlabel= {SNR [dB]},
xmajorgrids,
ymin=0.0001,
ymax=4,
ylabel={BER},
ymajorgrids,
legend entries={TI ZX MMDDT \cite{viveros2020_ICASSP}, proposed waveform,  random ZX transceiver \cite{deng2019bandlimited}, Golay ZX transceiver \cite{deng2019bandlimited} },
legend style={at={(0,0)},anchor=south  west,draw=black,fill=white,legend cell align=left,font=\tiny}
]
\addlegendimage{smooth,color=black, only marks, line width=0.75pt, style={solid,fill=black!50}, mark=diamond*, 
y filter/.code={\pgfmathparse{\pgfmathresult-0}\pgfmathresult}}
\addlegendimage{smooth,color=red, only marks, line width=0.75pt, style={solid,fill=red!50}, mark=*,
y filter/.code={\pgfmathparse{\pgfmathresult-0}\pgfmathresult}}
\addlegendimage{smooth,color=cyan, only marks, line width=0.75pt, style={solid,fill=cyan!50}, mark=square*,
y filter/.code={\pgfmathparse{\pgfmathresult-0}\pgfmathresult}}
\addlegendimage{smooth,color=green,only marks, line width=0.75pt, style={solid,fill=green!50}, mark=triangle*, 
y filter/.code={\pgfmathparse{\pgfmathresult-0}\pgfmathresult}}


\addplot+[solid, color=black, every mark/.append style={solid, fill=black!50}, mark=diamond*, mark repeat=1, line width=1.0pt]
  table[row sep=crcr]{%
-10	0.459263888888889\\
-8	0.440305555555556\\
-6	0.411416666666667\\
-4	0.381763888888889\\
-2	0.341277777777778\\
0	0.299152777777778\\
2	0.246069444444444\\
4	0.197222222222222\\
6	0.146472222222222\\
8	0.101680555555556\\
10	0.0615555555555556\\
12	0.0307916666666667\\
14	0.01225\\
16	0.00355555555555556\\
18	0.000791666666666667\\
20	8.33333333333333e-05\\
22	1.38888888888889e-05\\
24	0\\
26	0\\
28	0\\
30	0\\
  };

\addplot+[solid, color=red, every mark/.append style={solid, fill=red!50}, mark=*, mark repeat=1, line width=1.0pt]
  table[row sep=crcr]{%
-10	0.463776666666667\\
-8	0.444889166666667\\
-6	0.419326666666667\\
-4	0.384619166666667\\
-2	0.342468333333333\\
0	0.292285833333333\\
2	0.23956\\
4	0.189190833333333\\
6	0.1420225\\
8	0.0996891666666667\\
10	0.0627008333333333\\
12	0.0326841666666667\\
14	0.0129175\\
16	0.00342583333333333\\
18	0.000538333333333333\\
20	4.08333333333333e-05\\
22	2.5e-06\\
24	0\\
26	0\\
28	0\\
30	0\\
};
\addplot+[solid, color=green, every mark/.append style={solid, fill=green!50}, mark=triangle*, mark repeat=1, line width=1.0pt]
  table[row sep=crcr]{%
-10	0.382305555555556\\
-8	0.358513888888889\\
-6	0.322865740740741\\
-4	0.289509259259259\\
-2	0.251296296296296\\
0	0.21749537037037\\
2	0.184787037037037\\
4	0.154106481481481\\
6	0.124703703703704\\
8	0.0973148148148148\\
10	0.072287037037037\\
12	0.0524351851851852\\
14	0.0365231481481481\\
16	0.0237546296296296\\
18	0.0160185185185185\\
20	0.00923148148148148\\
22	0.00489814814814815\\
24	0.0024212962962963\\
26	0.00106018518518519\\
28	0.000412037037037037\\
30	0.000152777777777778\\
};
%
\addplot+[solid, color=cyan, every mark/.append style={solid, fill=cyan!50}, mark=square*, mark repeat=1, line width=1.0pt]
  table[row sep=crcr]{%
-10	0.455548611111111\\
-8	0.445993055555556\\
-6	0.433833333333333\\
-4	0.415256944444444\\
-2	0.398715277777778\\
0	0.375854166666667\\
2	0.350763888888889\\
4	0.316743055555556\\
6	0.279722222222222\\
8	0.233576388888889\\
10	0.189520833333333\\
12	0.146125\\
14	0.103972222222222\\
16	0.0628194444444444\\
18	0.0356944444444444\\
20	0.0210208333333333\\
22	0.0112986111111111\\
24	0.00513888888888889\\
26	0.00209722222222222\\
28	0.000631944444444444\\
30	0.000236111111111111\\
};

\addplot[color=black,dashed, mark=none, line width=0.75pt,
y filter/.code={\pgfmathparse{\pgfmathresult-0}\pgfmathresult}]
  table[row sep=crcr]{%
	1 50\\
};\label{iterC55}

\addplot[color=black, solid, mark=none, line width=0.75pt,
y filter/.code={\pgfmathparse{\pgfmathresult-0}\pgfmathresult}]
  table[row sep=crcr]{%
	1 50\\
};\label{iterC66}


\end{axis}

\end{tikzpicture}%
    \end{minipage}%
    \begin{minipage}[t]{0.33\textwidth}
      \centering
      \input{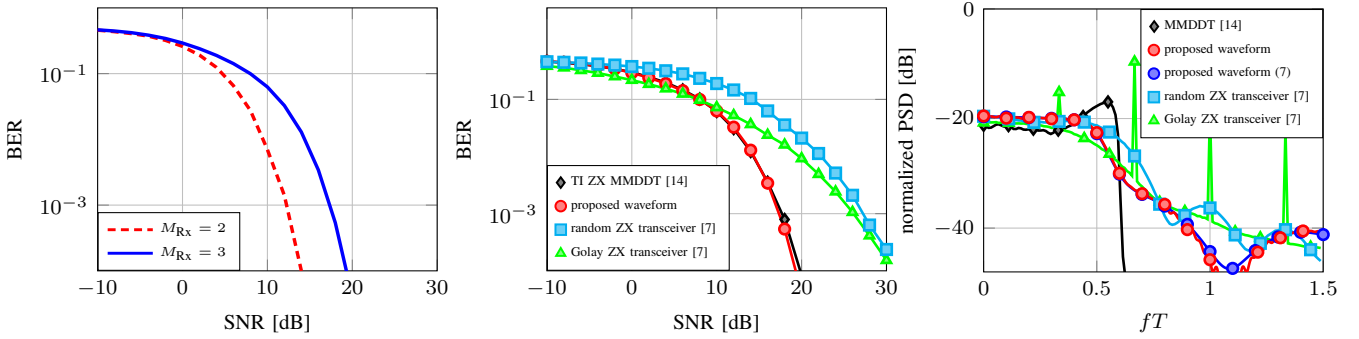}
    \end{minipage}
    \vspace{-4mm}
    \caption{Numerical evaluations. In (a) BER vs SNR for the proposed waveform. In (b) BER vs SNR for  $M_{\text{Rx}}= 3$ for all the considered methods.
    In (c) PSD for $M_{\text{Rx}}= 3$.\looseness-1\vspace{-5mm}}
    \label{fig:num_res} 
\end{figure*}

\section{Numerical Results}
\label{sec:numericalR}
This section presents numerical uncoded BER results and normalized PSD for the proposed TI ZX state machine waveform design with power containment factor $\eta = 0.95$. Moreover, the proposed technique results are compared with other methods from the literature, namely TI ZX MMDDT \cite{viveros2020_ICASSP} and ZX transceiver design \cite{deng2019bandlimited}. The channel considers $N_{\text{t}} = 8$ transmit antennas and $N_{\text{u}}= 2$ single antenna users for all the evaluated methods. The SNR is defined as follows
\begin{equation}
    \mathrm{SNR} = \frac{ E_0/(NT) }{N_0 B} = \frac{E_0}{N T N_0 2 f_{c}},
\end{equation}
where $N_0$ denotes the noise power spectral density. The  bandwidth $B$ is define as $B = 2f_{c}$, where the critical frequency is set to $f_c = 0.65/T$. The entries of the channel matrix $\boldsymbol{H}$ are i.i.d. with $\mathcal{C}\mathcal{N}(0,1)$.

The presented results for the TI ZX MMDDT method from \cite{viveros2020_ICASSP} considers  ${M_\mathrm{Rx}} = 3$ and the same data rate as for the proposed TI ZX  state machine waveform design with $g_{\text{Tx}}(t)$ as an RC filter and $g_{\text{Rx}}(t)$  as an RRC filter with roll-off factors  $\epsilon_{\text{Tx}} = \epsilon_{\text{Rx}} = 0.22$, with $f_{c} = (1+\epsilon_{\text{Tx}})/2T $. On the other hand, for the ZX transceiver design \cite{deng2019bandlimited}, ${M_\mathrm{Rx}} = 3$ is considered for the random and the Golay mapping methods. The truncation interval is set to  $\kappa = 3$ and the number of bits per subinterval $n = 2$, and at the receiver an integrate-and-dump-filter is considered  \cite{deng2019bandlimited}. Table~\ref{tab:num_res_parameters} summarizes the simulation parameters for the proposed TI ZX waveform design and other methods from the literature, where $I_\textrm{b}$ corresponds to the number of input bits per user and $O_\textrm{s}$ represents the number of samples after the mapping process.

The  optimal matrix $\boldsymbol{G}$ of positive coefficients   is shown in Table~\ref{tab:optCoeffMrx2} and Table~\ref{tab:optCoeffMrx3} for ${M_\mathrm{Rx}} = 2$ and  ${M_\mathrm{Rx}} = 3$, respectively, \textcolor{g}{where the normalization $m\frac{E_{\text{0}}}{2 N_{\text{tot}}} = 1$ is considered for the problem in \eqref{eq:OPproblem}}. The input sequences of symbols $\boldsymbol{x}$ are mapped onto the temporal transmit vector $\boldsymbol{s}_{g}$ considering the set of coefficients in Table~\ref{tab:optCoeffMrx2} and Table~\ref{tab:optCoeffMrx3}. The numerical BER results for the proposed TI ZX  state machine  waveform design are presented in Fig.~\ref{fig:num_res} (a) for ${M_\mathrm{Rx}} = 2$ and ${M_\mathrm{Rx}} = 3$. As expected, the BER for ${M_\mathrm{Rx}} = 2$ is lower than for ${M_\mathrm{Rx}} = 3$.
In Fig.~\ref{fig:num_res} (b) the BER is evaluated and compared with other methods form the literature for ${M_\mathrm{Rx}} = 3$. The TI ZX MMDDT \cite{viveros2020_ICASSP} and the proposed TI ZX  state machine  waveform design achieves approximately the same BER performance while the proposed  TI ZX  state machine waveform design has a lower computational complexity. In this context, the complexity order for the proposed state machine waveform design is dominated by the spatial ZF precoder whose complexity in Big O nation is given by $\mathcal{O}\left ( N_{\text{t}}^{3}\right )$. This is because the coefficients are optimized only once for any transmit sequence of symbols. On the other hand, the complexity order for the TI ZX MMDDT \cite{viveros2020_ICASSP} is given by $\mathcal{O}\left ( 2N_{\text{u}}(N_{\text{tot}})^{3.5} + N_{\text{t}}^{3}\right )$. However, note that  the proposed TI ZX state machine waveform design yields a low level of out-of-band-radiation as seen in Fig.~\ref{fig:num_res} (c).
Additionally, the proposed method is compared with the transceiver design from \cite{deng2019bandlimited}. The transceiver design method considers the nonuniform zero-crossing pattern with random and Golay mapping and power containment factor $\eta = 0.95$.

Simulation results are presented also in terms of the normalized PSD.  In Fig.~\ref{fig:num_res} (c) the analytical and numerical PSD are compared for the proposed TI ZX state machine waveform design with ${M_\mathrm{Rx}} = 3$. The analytical PSD is calculated with \eqref{psdEquation} considering the autocorrelation function in \eqref{autocorrAverage}. In Fig.~\ref{fig:num_res} (c), the normalized PSD of the proposed waveform design is also compared with the normalized PSD of the methods from the literature which is calculated by 
\begin{align}
\textrm{PSD}_{\textrm{dB}} = 10\textrm{log}_{10}\left [ O_{\text{s}}^{(-1)}\mathrm{E}  \{ \left | F_{i} \right |^2 \}\right ]\text{,}
\end{align}
where $F_{i}$ is the discrete Fourier transform of the normalized temporal transmit signal per user.

\section{Conclusions}
\label{sec:Fconclusions}
In this study, we have developed a TI ZX state machine waveform based on the novel TI ZX modulation for multi-user MIMO downlink systems, with 1-bit quantization and oversampling. The waveform design considers the optimization of a set of coefficients which conveys the information into the time-instances of zero-crossings. The optimization is performed considering the power containment bandwidth and the maximization of the minimum distance to the decision threshold. The simulation results were compared with methods from the literature which employ techniques based on zero-crossings. The BER performance is favorable for the proposed method which achieves a comparable BER result as the TI ZX MMDDT \cite{viveros2020_ICASSP} method but with significantly lower computational complexity. 
\begin{table}[!t]
    \caption{Optimal set $\boldsymbol{G}$ for ${M_\mathrm{Rx}} = 2$}
    \label{tab:optCoeffMrx2} 
    \begin{center}
    \scalebox{0.7}{
\begin{tabular}{|ccccc|}
\hline
\multicolumn{5}{|c|}{$\boldsymbol{G}$}                                                 \\ \hline
\multicolumn{1}{|c|}{$\boldsymbol{g}_{1}$} & $0.2719$, & $0.3751$, & $0.3715$, & $0.2378$ \\ \hline
\multicolumn{1}{|c|}{$\boldsymbol{g}_{2}$} & $0.2081$, & $0.2129$, & $0.1$,    & $0.1$    \\ \hline
\multicolumn{1}{|c|}{$\boldsymbol{g}_{3}$} & $0.1719$, & $0.1$,    & $0.1$,    & $0.1440$ \\ \hline
\multicolumn{1}{|c|}{$\boldsymbol{g}_{4}$} & $0.1$,    & $0.1$,    & $0.1832$, & $0.1572$ \\ \hline
\multicolumn{1}{|c|}{$\boldsymbol{g}_{5}$} & $0.1$,    & $0.1$,    & $0.1$,    & $0.1$    \\ \hline
\multicolumn{1}{|c|}{$\boldsymbol{g}_{6}$} & $0.1$,    & $0.2030$, & $0.1$,    & $0.1$    \\ \hline
\multicolumn{1}{|c|}{$\boldsymbol{g}_{7}$} & $0.1$,    & $0.2507$, & $0.2551$, & $0.1655$ \\ \hline
\multicolumn{1}{|c|}{$\boldsymbol{g}_{8}$} & $0.1$,    & $0.1$,    & $0.1$,    & $0.1647$ \\ \hline
\end{tabular}}
    \end{center}
    \vspace{-2mm}
\end{table}
\begin{table}[!t]
    \caption{Optimal set $\boldsymbol{G}$ for ${M_\mathrm{Rx}} = 3$}
    \label{tab:optCoeffMrx3} 
    \begin{center}
    \scalebox{0.7}{
\begin{tabular}{|cccc|}
\hline
\multicolumn{4}{|c|}{$\boldsymbol{G}$}                                        \\ \hline
\multicolumn{1}{|c|}{$\boldsymbol{g}_{1}$} & $0.4566$, & $0.4809$, & $0.4006$ \\ \hline
\multicolumn{1}{|c|}{$\boldsymbol{g}_{2}$} & $0.2631$, & $0.1$,    & $0.1014$ \\ \hline
\multicolumn{1}{|c|}{$\boldsymbol{g}_{3}$} & $0.1334$, & $0.1$,    & $0.2312$ \\ \hline
\multicolumn{1}{|c|}{$\boldsymbol{g}_{4}$} & $0.1$     & $0.2875$  & $0.3692$ \\ \hline
\end{tabular}}
    \end{center}
    \vspace{-6mm}
\end{table}

\section*{\bfseries{Acknowledgements}}
This work has been supported by FAPERJ, the ELIOT ANR-18-CE40-0030 and FAPESP 2018/12579-7 project.

\bibliographystyle{IEEEbib}
\bibliography{ref}

\end{document}